\documentclass[aps,prl,twocolumn,showpacs,superscriptaddress,amsmath,
               nobalancelastpage]{revtex4}

\usepackage{graphicx}
\usepackage[tight]{subfigure}
\usepackage{bbm}

\newcommand{\identity}{\mathbbm{1}}
\newcommand{\bra}[1]{\mathinner{\langle{#1}|}}
\newcommand{\ket}[1]{\mathinner{|{#1}\rangle}}
\newcommand{\braket}[2]{\mathinner{\langle{#1}|{#2}\rangle}}

\newcommand{\ketbra}[3][]{\ket{#2}_{#1}\hspace{-0.8mm}\bra{#3}}

\newcommand{\proj}[2][]{\ketbra[#1]{#2}{#2}}

\newcommand{\Prj}{\Pi}
\newcommand{\abs}[1]{|#1|}

\newcommand{\matnorm}[1]{||#1||}
\newcommand{\Matnorm}[1]{\left\Vert #1 \right\Vert}
\DeclareMathOperator{\tr}{tr}
\DeclareMathOperator{\diag}{diag}

\newcommand{\Ueff}{U_\mathrm{eff}}
\newcommand{\sep}[2]{$#1$--$#2$}

\begin{document}

\title{Separable states can be used to distribute entanglement}

\date{February 21, 2003}

\author{T. S. Cubitt}
\author{F. Verstraete}
\affiliation{Max Plank Institut f\"ur Quantenoptik, Hans--Kopfermann Str.\ 1,
  D-85748 Garching, Germany}
\author{W. D\"ur}
\affiliation{Sektion Physik, Ludwig-Maximilians-Universit\"at M\"unchen,
  Theresienstr.\ 37, D-80333 M\"unchen, Germany}
\author{J.I. Cirac}
\affiliation{Max Plank Institut f\"ur Quantenoptik, Hans--Kopfermann Str.\ 1,
  D-85748 Garching, Germany}

\pacs{03.67.Mn, 03.67.-a}

\begin{abstract}
  We show that no entanglement is necessary to distribute entanglement; that
  is, two distant particles can be entangled by sending a third particle that
  is never entangled with the other two. Similarly, two particles can become
  entangled by continuous interaction with a highly mixed mediating particle
  that never itself becomes entangled. We also consider analogous properties of
  completely positive maps, in which the composition of two separable maps can
  create entanglement.
\end{abstract}

\maketitle


Einstein associated entanglement with ``spooky action-at-a-distance'', a
strange quantum effect that did not tally with his ideas of how the universe
ought to work~\cite{EPR35}. Modern quantum information theory, however, takes
a different view: entanglement is a physical quantity. And like other physical
quantities, it can be used as a resource. The major successes in the field
have come from asking: ``what new possibilities arise when entanglement is
available?''. The power of this quantum resource became especially apparent
after the discovery of quantum teleportation~\cite{BBC+95}, which showed that,
if entanglement and classical communication are available, global quantum
operations can be implemented locally. Since then, huge progress has been made
in describing the way entanglement can be distributed and manipulated among
separated parties. As for other physical quantities, conservation laws have
been formulated dictating e.g.\ that the amount of entanglement can not be
increased by local operations and classical communication (LOCC)~\cite{BDSW96}.
Specifically, this means entanglement can only be created by an interaction
between particles.

\begin{figure}[htb]
  \subfigure[]{%
    \label{subfig:discrete1}\includegraphics{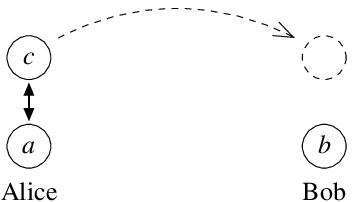}%
    }%
  \hspace{30pt}%
  \subfigure[]{%
    \label{subfig:discrete2}\includegraphics{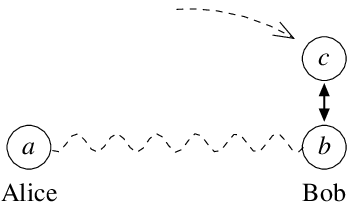}%
    }
  \caption{Alice and Bob each have a particle they wish to entangle
    with the other. \subref{subfig:discrete1}~Alice interacts an ancilla $c$
    with her particle $a$, sends $c$ to Bob, \subref{subfig:discrete2}~who
    interacts $c$ with his particle $b$. At the end they share some
    entanglement. Surprisingly, $c$ does not have to become entangled
    with $a$ and $b$.
    \label{fig:discrete}
    }
\end{figure}

In this Letter, we investigate more closely the conditions required to
entangle two distant particles. Though there is no unique way to quantify
entanglement for mixed states, the usual definition of an entangled state --
as one that can not be created by LOCC -- \emph{is} unambiguous. To create
entanglement, then, a global quantum operation is necessary. For separated
particles, this must be carried out by sending a mediating particle between
them (see Fig.~\ref{fig:discrete}), i.e.\ by communication via a quantum
channel. Note that fundamentally this is the only way entanglement is
created~\footnote{Experiments often produce entanglement via indirect
  interactions. In ion-traps, electronic states of the ions interact via
  phonon modes; in cavities, atoms are entangled via interactions with photons
  in the cavity. (See e.g.~\cite{FortPhys}.)}, as all interactions in nature
occur via mediating gauge bosons. It is clear that the particles can be
entangled if the mediating (or `ancilla') particle becomes entangled: entangle
the first particle with the ancilla, send the latter through the channel, and
swap it with the second particle. We can picture the entanglement being sent
\emph{through} the channel. One would expect that the ancilla
\emph{necessarily} becomes entangled, in any scheme.  Surprisingly, we will
prove that this intuitive picture is false, and that two particles can become
entangled without the ancilla \emph{ever} becoming entangled. (Note that this
does not imply entanglement can be created by LOCC since a quantum channel was
used.)

\begin{figure}[htb]
  \includegraphics{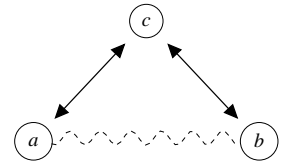}
  \caption{Particles $a$ and $b$ interact continuously via
    a mediating particle $c$. For certain initial states, $c$ remains
    separable from $a$ and $b$ \emph{at all times} during the evolution of the
    system, yet at the end $a$ and $b$ are entangled.
    \label{fig:continuous}
    }
\end{figure}

We demonstrate this fact in two different ways. First we consider a process in
which two particles interact continuously with an ancilla
(Fig.~\ref{fig:continuous}). We prove that if it is possible to entangle the
particles while leaving the ancilla separable at all times, it is possible to
turn this into a discretized scheme in which an ancilla is sent between Alice
and Bob a number of times. Inspired by quantum optical systems, we give an
example of such a continuous process in which the ancilla remains separable.

As a second demonstration, we give an explicit discrete procedure in which an
ancilla is sent from Alice to Bob, and the interactions are described by local
unitary operations: a single use of a quantum channel is already sufficient to
distribute entanglement \emph{without} sending entanglement. By analyzing
bipartite partitions of the tripartite system, we explain why the process
works. Finally, we translate this result describing properties of states into
a result describing properties of completely positive maps (CPM): we give an
example of two maps that individually cannot create entanglement from
initially separable states, nevertheless their composition can.

First consider the continuous case. As a model of an interaction mediated by
another particle, consider a system of two particles `$a$' and `$b$' both
interacting with the same ancilla `$c$' (Fig.~\ref{fig:continuous}) but not
directly with each other. If we restrict ourselves to pure states, the effect
presented above is impossible: to entangle $a$ and $b$, $c$ must become
entangled with respect to the partition \sep{(ab)}{c}~\footnote{We say a state
  $\rho_{abc}$ is separable with respect to the partition \sep{(ab)}{c} iff
  $\rho_{abc}=\sum_k \proj[ab]{\varphi_k} \otimes \proj{c_k}$, and is
  entangled otherwise.}. This is easily proven by looking at the infinitesimal
change of a pure separable state $\ket{a}\ket{b}\ket{c}$ under the action of a
Hamiltonian of the form $H_{AC}\otimes \identity_B + \identity_A\otimes
H_{BC}$. The condition that the ancilla remains separable is given by
\begin{multline}\label{eq:pure}
  \bigl( \identity + \delta t(H_{AC}+H_{BC}) \bigr) \ket{a}\ket{b}\ket{c}\\
  = \bigl( \ket{a}\ket{b} + \delta t\ket{\psi_{ab}} \bigr) \bigl(\ket{c} +
  \delta t\ket{\psi_c}\bigr)
\end{multline}
where $\ket{\psi_{ab}}$ and $\ket{\psi_c}$ are unnormalized states. It is
readily seen that the state $\ket{a}\ket{b} + \delta t\ket{\psi_{ab}}$ is
entangled to first order in $\delta t$ if and only if $\ket{\psi_{ab}}$ has a
non-vanishing component $\bra{a^\bot}\braket{b^\bot}{\psi_{ab}}$ for some
$\ket{a^\bot},\ket{b^\bot}$ orthogonal to $\ket{a},\ket{b}$. However,
multiplying equation~\eqref{eq:pure} by $\bra{a^\bot}\bra{b^\bot}$ shows this
is not the case. This is in complete accordance with the intuition that
entanglement can only be transmitted if the mediating particle itself becomes
entangled.

Let us however investigate how the amount of entanglement of the ancilla is
related to the amount of entanglement that can be transmitted. Physically, the
ancilla has a spectrum of energy levels corresponding to eigenstates of its
Hamiltonian. Imagine that interactions with particles $a$ and $b$ couple
eigenstates of the ancilla, and the interactions are weak so that transitions
between energy levels are virtual. The most significant processes in the
evolution of the system are then second order virtual transitions of $c$ from
an energy level back to itself, accompanied by an interaction between
particles $a$ and $b$. If the ancilla starts in an energy eigenstate, the
virtual transition will leave it in the same eigenstate, not entangled with
$a$ or $b$. It \emph{will} become entangled with $a$ or $b$ due to less
significant higher order processes, but this entanglement will be weak.
Particles $a$ and $b$, however, can become strongly entangled if the
interaction lasts a long time, and we have a situation close to the advertised
effect: $a$ and $b$ can become entangled while the ancilla remains
\emph{almost} separable. But assume now that there is noise in the system,
which can be modeled by mixing the quantum states with the maximally mixed
state. Mixing a barely entangled state with an amount of noise proportional to
the entanglement destroys the entanglement~\cite{VT99,BCJ+99}. We can
therefore add a sufficiently large weight of the maximally mixed state to
remove the weak entanglement with the ancilla, but not so large that it
destroys the strong entanglement between $a$ and $b$. (This is related to the
results that entanglement can be created by interaction with a common heat
bath~\cite{Bra02}). If mixed states are considered, it therefore seems to be
possible to obtain the desired effect in the continuous regime.

But this implies the seemingly stronger result that a separable ancilla
bouncing back and forth between two distant particles (see
Fig.~\ref{fig:discrete}) can distribute (distillable) entanglement. A
continuous interaction whose evolution operator is of the form $e^{-i(H_A +
  H_B)t}$ can be discretized, e.g.\ using the Trotter expansion: $e^{A+B} =
\lim_{n\rightarrow\infty}(e^{A/n}e^{B/n})^n$. Note that the discretized
evolution can be made arbitrary close to the original one by choosing a large
(but finite) $n$. This strange effect forces us to completely abandon the
physical picture of sending entanglement \emph{through} a quantum channel.
E.g.\ this implies that the Ekert protocol in quantum
cryptography~\cite{Eke91} could be implemented by sending separable states.

To demonstrate these ideas, let us consider a specific example of the
continuous case, in which $a$ and $b$ are qubits and the ancilla $c$ is a
qutrit with eigenstates $\ket{0}$, $\ket{1}$ and $\ket{2}$. Motivated by
ion-trap quantum computation experiments (see e.g.\ Ref.~\cite{WBB+02}), in
which vibrational modes of the trapped ions play the role of the
ancilla~\cite{Cir95,SM99}, or by cavity-QED experiments~\cite{OBA+01} in which
the ancilla is the cavity mode, we consider the Hamiltonian
\begin{equation*}
  H = \hat{c}^\dagger \hat{c} \otimes\identity + \frac{\epsilon}{2}
    (\hat{c}+\hat{c}^\dagger)\otimes\left(\sigma_x^a + \sigma_x^b\right),
\end{equation*}
where $\epsilon$ is small and characterizes the strength of the interaction,
and $\hat{c}^\dagger$ is the ``creation'' operator in the truncated Hilbert
space of the ancilla: $\hat{c} = \ketbra[c]{0}{1} + \sqrt{2}\ketbra[c]{1}{2}$,
so $\hat{c}^\dagger \hat{c} = \proj[c]{1} + 2\proj[c]{2} = H_0$. The
interactions couple eigenstates of the ancilla, as described above. As we
are interested in the leading order evolution (and how much the true evolution
deviates from this), we turn to perturbation theory. The Hamiltonian operates
in four invariant subspaces: $H = H_+^c\otimes\Prj_{++}^{ab} +
H_-^c\otimes\Prj_{--}^{ab} + H_0^c\otimes\left(\Prj_{+-}^{ab} +
  \Prj_{-+}^{ab}\right)$, where $H_\pm = \hat{c}^\dagger \hat{c} \pm
\epsilon(\hat{c} + \hat{c}^\dagger)$, and the $\Prj$'s denote projectors whose
subscripts refer to the $\ket{+},\ket{-}$ qubit states with $\ket{\pm} =
1/\sqrt{2}(\ket{0}\pm\ket{1})$.  The evolution operator can then be written as
a direct sum of operators on these spaces: $U = e^{-iH_+t} \oplus e^{-iH_0t}
\oplus e^{-iH_0t} \oplus e^{-iH_-t}$. Expanding $H_\pm$ using standard
perturbation theory, we have $e^{-iH_\pm t} = e^{-iDt} + O(\epsilon^4 t) +
O(\epsilon)$, where \mbox{$D = \diag(-\epsilon^2, 1-\epsilon^2,
  2+2\epsilon^2)$} is the matrix of eigenvalues of $H_\pm$ approximated to
leading order (accurate to $3^{rd}$ order since all odd order terms are zero).
The $O(\epsilon^4t)$ term arises from higher order perturbations of the
eigenvalues, the $O(\epsilon)$ term from perturbation of the eigenstates. We
can then approximate the evolution operator:
\begin{equation}\label{eq:U_approx}
  U = \Ueff + O(\epsilon^4t) + O(\epsilon),
\end{equation}
where $\Ueff = e^{-iDt} \oplus e^{-iH_0t} \oplus e^{-iH_0t} \oplus e^{-iDt}$.
In the limit of small $\epsilon$, the evolution of the system is ``close'' to
evolution under $\Ueff$. How do states evolve under $\Ueff$? If we start $c$
in an eigenstate of $\Ueff$ ($\ket{0}$, $\ket{1}$ or $\ket{2}$), its state
remains unchanged up to a global phase, and can not become entangled with $a$
and $b$. If $a$ and $b$ start in a superposition of eigenstates of $\Ueff$,
the $\ket{++}$ and $\ket{--}$ portions acquire a phase difference relative to
the $\ket{+-}$ and $\ket{-+}$ portions ($e^{-2i\epsilon^2 t}$ if $c$ is in the
$\ket{2}$ state, $e^{i\epsilon^2t}$ otherwise, recalling the expressions for
$D$ and $H_0$). Particles $a$ and $b$ can therefore become highly entangled
for times of order $1/\epsilon^2$, even though the ancilla remains separable
at all times. The true evolution under $U$ deviates from that under $\Ueff$
due to the higher order terms in Eq.~\eqref{eq:U_approx}, and \emph{will}
entangle the ancilla if the system starts in a pure state. To show that a
small amount of mixing is sufficient to remove this entanglement while leaving
$a$ and $b$ highly entangled, we must bound this deviation at times $t\sim
1/\epsilon^2$ to $O(\epsilon)$ (that is, the deviation between what we
\emph{would} get if the system evolved under $\Ueff$, and what we \emph{do}
get when the system evolves under $U$). This means bounding the matrix norm
$\matnorm{ e^{-iH_\pm t} - e^{-iDt}}$. To achieve this, we make use of the
matrix of eigenvectors of $H_\pm$ approximated to third order in $\epsilon$,
denoted $X$. Then
\begin{multline*}
  \Matnorm{e^{-iH_\pm t} - e^{-iDt}}
    \le \Matnorm{e^{-iH_\pm t} - Xe^{-iDt}X^{-1}}\\
  + \Matnorm{Xe^{-iDt}X^{-1} - e^{-iDt}}.
\end{multline*}
The first norm on the right hand side ($\mathcal{N}_1$) is governed by higher
order perturbations of the eigenvalues, and can be bounded to $O(\epsilon^4
t)$ by standard results in linear algebra~\cite[\S2.3.4 \& \S11.3.2]{GL96},
giving
\begin{equation*}
  \mathcal{N}_1 \le \Matnorm{X^{-1}}^2 \Matnorm{X\vphantom{^1}}
    \Matnorm{H_\pm X - XD\vphantom{^1}} t.
\end{equation*}  
The second norm ($\mathcal{N}_2$) is governed by perturbation of the
eigenvectors. It can be bounded to $O(\epsilon)$ by expanding the exponentials
as power series, and making use of the fact that a matrix satisfies its own
characteristic equation to reduce the expansions to finite order polynomials,
whose coefficients are functions of the eigenvalues $d_{1,\dots,3}$ of $D$ and
time $t$. The time dependence is removed by taking maxima of these
coefficients over all times, giving
\begin{multline*}
  \mathcal{N}_2 \le \matnorm{X^{-1}} \cdot \frac{2}{\abs{\Delta}}
  \Big( \left(d_3^2-d_1^2\right)\matnorm{XD-DX}\\
  + \left(d_3-d_1\right) \matnorm{XD^2-D^2X} \Big),
\end{multline*}
where $\Delta$ is the determinant of the Vandermonde matrix formed by
$d_{1,\dots,3}$.

We can now investigate the evolution of the initial state
\begin{equation*}
  \rho = \frac{1}{(1+\alpha)^2}
    \left(\proj[a]{0} + \alpha\tfrac{1}{2}\identity_a\right)
    \otimes\left(\proj[b]{0} + \alpha\tfrac{1}{2}\identity_b\right)
    \otimes\tfrac{1}{3}\identity_c
\end{equation*}
which is not only separable, but also classically uncorrelated. The ancilla is
in a mixture of eigenstates of $\Ueff$, so will not become entangled to
leading order (i.e.\ by $\Ueff$), whereas $a$ and $b$ are in a superposition
of the eigenstates of $\Ueff$ ($\ket{+},\ket{-}$) and do become
entangled, as discussed previously. Using the bounds $\mathcal{N}_{1,2}$, we
can bound the contribution of the higher order terms to the the evolved
density matrix. This allows us to derive an expression for the minimum amount
of mixing $\alpha$ that is required, for a given $\epsilon$, to ensure any
entanglement with the ancilla is removed. We derive a second inequality
involving $\alpha$ and $\epsilon$ by considering the maximum amount of mixing
that does not destroy the entanglement between $a$ and $b$~\cite{VV02}. We can
then calculate a bound on the $\epsilon$ that allow the two expressions to be
satisfied simultaneously, i.e.\ the interaction strengths for which there is
some $\alpha$ that removes all entanglement with the ancilla but leaves
entanglement between $a$ and $b$. We obtain the condition $\epsilon < 1/7000$.
We have therefore proven that the desired effect is physical. Numerically we
find that this bound is much too strict, and $\epsilon \lesssim 1/10$ is
already sufficient.

Let us now move to the discrete case (see Fig.~\ref{fig:discrete}). Alice and
Bob start by preparing a (classically correlated) separable state in step~1,
where Alice has particles $a$ and $c$, Bob has particle $b$. In step~2, Alice
applies an operation on her two particles $a$ and $c$, then sends particle $c$
to Bob. Finally, in step~3 Bob applies an operation on $b$ and $c$, resulting
in a state that contains (distillable) entanglement between $a$ and $b$
(tracing out $c$), even though $c$ has remained separable from $(ab)$
throughout.

This quantum information picture gives us more insight into the basic
principles underlying this (bipartite) effect, in which entanglement
properties of tripartite systems turn out to play a key role. In tripartite
systems, the entanglement properties of all bipartite partitions are
independent~\cite{Dur99}. For example, there exist mixed states in which $c$
is separable from $(ab)$, $b$ is separable from $(ac)$, yet $a$ is entangled
with $(bc)$. Indeed, step~2 creates a state with precisely these properties.
Step~3 then entangles $b$ with $(ac)$ without changing the entanglement
properties of the other bipartite partitions. At each step, $c$ remains
separable from the rest of the system.

We now give an explicit example in which all particles are qubits,
demonstrating the above process. Step~1: Alice and Bob prepare the (manifestly
separable) classically correlated state
\begin{equation*}
  \rho_{abc} = \frac{1}{6}\sum_{k=0}^3 \proj{\Psi_k,\Psi_{-k},0}
  + \sum_{i=0}^1 \frac{1}{6}\proj{i,i,1}
\end{equation*}
where $\ket{\Psi_k} = 1/\sqrt{2}\left( \ket{0} + e^{ik\pi/2}\ket{1} \right)$.
Step~2: Alice applies a CNOT operation on particles $a$ and $c$ (where $a$ is
the control qubit), producing the state
\begin{equation}
  \sigma_{abc} = \frac{1}{3} \proj{\Psi_{GHZ}}
  + \sum_{i,j,k=0}^1 \beta_{ijk}\Prj_{ijk}
  \label{eq:GHZ}
\end{equation}
where $\ket{\Psi_{GHZ}}_{abc} = 1/\sqrt{2}\left(\ket{000}+\ket{111}\right)$,
$\Prj_{ijk} = \proj{ijk}$, and all $\beta$'s are $0$ apart from $\beta_{001} =
\beta_{010} = \beta_{101} = \beta_{110} = 1/6$. This state is invariant under
a permutation of the qubits $b$ and $c$, and must therefore be separable with
respect to the \sep{b}{(ac)} and \sep{c}{(ab)} partitions. Step~3: Bob applies
a CNOT on $b$ and $c$ (with $b$ as the control qubit), resulting in the state
\begin{equation*}
  \tau_{abc} = \frac{1}{3}\proj[ab]{\phi^+}\otimes\proj[c]{0} +
  \frac{2}{3}\identity_{ab}\otimes\proj[c]{1}
\end{equation*}
where $\ket{\phi^+} = 1/\sqrt{2}\left(\ket{00} + \ket{11}\right)$ is maximally
entangled. The ancilla $c$ clearly remains separable with respect the rest of
the system, but the state now contains entanglement between $a$ and $b$.

Bob can extract this entanglement in a number of ways. Measuring $c$ in the
computational basis, he can extract a maximally entangled state of $ab$ with
probability $1/3$. Alternatively, if a deterministic effect is required, he
can apply a local completely positive map (CPM) to particles $b$ and $c$,
defined by $\mathcal{E}_{bc}(\rho) = \sum_j
O_{bc}^{(j)}\rho\,O_{bc}^{(j)\dagger}$ with Kraus operators $O_{bc}^{(1)} =
\identity_b\otimes\proj[c]{0}$, $O_{bc}^{(2)} =
\proj[b]{0}\otimes\proj[c]{1}$, and $O_{bc}^{(3)} =
\ketbra[b]{0}{1}\otimes\proj[c]{1}$, satisfying $\sum_j O^{(j)\dagger} O^{(j)}
= \identity$. Throwing away the ancilla leaves the state $\rho_{ab} =
\tr_c\left(\mathcal{E}_{bc}(\tau_{abc})\right) = 1/3\proj{\phi^+} +
1/3\,\Prj_{00} + 1/3\,\Prj_{10}$, where $\Prj_{ij} = \proj{ij}$. This has
non-positive partial transpose, so must be (distillable)
entangled~\cite{Per96,HHH97}. Thus Alice and Bob achieve the announced effect.

An effect in the same spirit as the examples presented for (mixed) states can
also be found for quantum operations, i.e.\ for trace-preserving completely
positive maps (TPCPM). The result is a direct application of the duality
between non-local maps and states~\cite{CDK01} to the present puzzle. We will
prove that a composition of two \emph{separable} maps can be entangling. More
precisely, we consider a first (nonlocal) map acting on three qubits $abc$
that cannot create entanglement with respect to the bipartite partitions
\sep{b}{(ac)} and \sep{c}{(ab)}, and a second map that cannot create
entanglement with respect to \sep{a}{(bc)} and \sep{c}{(ab)}. Obviously,
neither of the maps can create entanglement between $a$ and $b$.
Interestingly, however, it turns out that the sequential application of these
TPCPM's can create (distillable) entanglement between $a$ and $b$.

Let us consider the TPCPM $\mathcal{E}_1$ with Kraus operators
\begin{equation}\label{eq:kraus}
  \begin{gathered}
    A_1=\proj{000}+\proj{111},\\
    \begin{aligned}
      A_2&=\proj{001}, &A_3&=\proj{010},        &A_4&=\proj{101},\\
      A_5&=\proj{110}, &A_6&=\ketbra{000}{011}, &A_7&=\ketbra{111}{100}.
    \end{aligned}
  \end{gathered}
\end{equation}
Using the duality between maps and states developed in~\cite{CDK01}, one can
easily show that this map cannot create entanglement in the partitions
\sep{b}{(ac)} and \sep{c}{(ab)}: the CPM corresponding to Kraus operators
$\{A_1\dots A_5\}$ of Eq.~\eqref{eq:kraus} gives rise to a dual state that is
exactly of the form of Eq.~\eqref{eq:GHZ}, which is separable in the stated
partitions. As this state can be used to implement the map probabilistically,
the map itself cannot be entangling in \sep{b}{(ac)} or in \sep{c}{(ab)}. For
the second map, we define the TPCPM $\mathcal{E}_2$ with Kraus operators as
for $\mathcal{E}_1$, but with the roles of $a$ and $b$ interchanged. This
cannot create entanglement in the \sep{a}{(bc)} and \sep{c}{(ab)} partitions.

The map $\mathcal{E}$ obtained by composing both maps
$\mathcal{E}=\mathcal{E}_2 \circ \mathcal{E}_1$ has Kraus operators $C_1=A_1$,
$C_2=A_2$, $C_3=\ketbra{000}{010}$, $C_4=A_6$, $C_5=A_7$,
$C_6=\ketbra{111}{101}$, $C_7=A_5$. One can readily prove that this map cannot
entangle states in the partition \sep{c}{(ab)}, but is entangling in the
partitions \sep{a}{(bc)} and \sep{b}{(ac)}. It immediately follows that the
composition map can create (distillable) entanglement between $a$ and $b$ out
of a separable state. For example, an entangled state $\rho_{AB}$ is obtained
by acting with $\mathcal{E}$ on the separable state $\ket{+++}$, when the map
is followed by a local measurement of $c$ in the $\ket{+},\ket{-}$ basis.

In this letter we have investigated the conditions required to entangle two
distant particles. We have shown that continuous interaction with a highly
mixed ancilla can create entanglement \emph{without} entangling the ancilla.
We have also shown that entanglement can be created by sending a
\emph{separable} ancilla, and that an analogous effect exists for quantum
operations. These results easily generalize to multi-party systems: many
particles interacting with a common ancilla, or an ancilla sent between each
party in turn. Thus no entanglement is required to create entanglement,
forcing us to abandon the picture of entanglement being sent \emph{through} a
quantum channel.

\begin{acknowledgments}
  We thank M. Lewenstein, J. M. Raimond, W. Zurek, and S. Popescu for valuable
  discussions. This work was supported in part by the E.C. (RESQ
  IST-2001-37559, HPMF-CT-2001-01209 (W.D.)) and the Kompetenznetzwerk
  "Quanteninformationsverarbeitung" der Bayerischen Staatsregierung.
\end{acknowledgments}

\bibliography{E-free-E.bib}

\begin{thebibliography}{18}
\expandafter\ifx\csname natexlab\endcsname\relax\def\natexlab#1{#1}\fi
\expandafter\ifx\csname bibnamefont\endcsname\relax
  \def\bibnamefont#1{#1}\fi
\expandafter\ifx\csname bibfnamefont\endcsname\relax
  \def\bibfnamefont#1{#1}\fi
\expandafter\ifx\csname citenamefont\endcsname\relax
  \def\citenamefont#1{#1}\fi
\expandafter\ifx\csname url\endcsname\relax
  \def\url#1{\texttt{#1}}\fi
\expandafter\ifx\csname urlprefix\endcsname\relax\def\urlprefix{URL }\fi
\providecommand{\bibinfo}[2]{#2}
\providecommand{\eprint}[2][]{\url{#2}}

\bibitem[{\citenamefont{Einstein et~al.}(1935)\citenamefont{Einstein, Podolsky,
  and Rosen}}]{EPR35}
\bibinfo{author}{\bibfnamefont{A.}~\bibnamefont{Einstein}},
  \bibinfo{author}{\bibfnamefont{B.}~\bibnamefont{Podolsky}}, \bibnamefont{and}
  \bibinfo{author}{\bibfnamefont{N.}~\bibnamefont{Rosen}},
  \bibinfo{journal}{Physical Review} \textbf{\bibinfo{volume}{47}},
  \bibinfo{pages}{777} (\bibinfo{year}{1935}).

\bibitem[{\citenamefont{Bennett et~al.}(1993)}]{BBC+95}
\bibinfo{author}{\bibfnamefont{C.~H.} \bibnamefont{Bennett}}
  \bibnamefont{et~al.}, \bibinfo{journal}{Phys.\ Rev.\ Lett.}
  \textbf{\bibinfo{volume}{70}}, \bibinfo{pages}{1895} (\bibinfo{year}{1993}).

\bibitem[{\citenamefont{Bennet et~al.}(1996)}]{BDSW96}
\bibinfo{author}{\bibfnamefont{C.~H.} \bibnamefont{Bennet}}
  \bibnamefont{et~al.}, \bibinfo{journal}{Phys.\ Rev.\ A.}
  \textbf{\bibinfo{volume}{54}}, \bibinfo{pages}{3824} (\bibinfo{year}{1996}).

\bibitem[{\citenamefont{Vidal and Tarrach}(1999)}]{VT99}
\bibinfo{author}{\bibfnamefont{G.}~\bibnamefont{Vidal}} \bibnamefont{and}
  \bibinfo{author}{\bibfnamefont{R.}~\bibnamefont{Tarrach}},
  \bibinfo{journal}{Phys.\ Rev.\ A.} \textbf{\bibinfo{volume}{59}},
  \bibinfo{pages}{141} (\bibinfo{year}{1999}).

\bibitem[{\citenamefont{Braunstein et~al.}(1999)}]{BCJ+99}
\bibinfo{author}{\bibfnamefont{S.~L.} \bibnamefont{Braunstein}}
  \bibnamefont{et~al.}, \bibinfo{journal}{Phys.\ Rev.\ Lett.}
  \textbf{\bibinfo{volume}{83}}, \bibinfo{pages}{1054} (\bibinfo{year}{1999}).

\bibitem[{\citenamefont{Braun}(2002)}]{Bra02}
\bibinfo{author}{\bibfnamefont{D.}~\bibnamefont{Braun}},
  \bibinfo{journal}{Phys.\ Rev.\ Lett.} \textbf{\bibinfo{volume}{89}},
  \bibinfo{pages}{277901} (\bibinfo{year}{2002}).

\bibitem[{\citenamefont{Ekert}(1991)}]{Eke91}
\bibinfo{author}{\bibfnamefont{A.}~\bibnamefont{Ekert}},
  \bibinfo{journal}{Phys.\ Rev.\ Lett.} \textbf{\bibinfo{volume}{67}},
  \bibinfo{pages}{661} (\bibinfo{year}{1991}).

\bibitem[{\citenamefont{Wineland et~al.}(2002)}]{WBB+02}
\bibinfo{author}{\bibfnamefont{D.~J.} \bibnamefont{Wineland}}
  \bibnamefont{et~al.}, \bibinfo{journal}{quant-ph/0212079}
  (\bibinfo{year}{2002}).

\bibitem[{\citenamefont{Cirac and Zoller}(1995)}]{Cir95}
\bibinfo{author}{\bibfnamefont{J.~I.} \bibnamefont{Cirac}} \bibnamefont{and}
  \bibinfo{author}{\bibfnamefont{P.}~\bibnamefont{Zoller}},
  \bibinfo{journal}{Phys.\ Rev.\ Lett.} \textbf{\bibinfo{volume}{74}},
  \bibinfo{pages}{4091} (\bibinfo{year}{1995}).

\bibitem[{\citenamefont{S{\"o}rensen and M{\o}lmer}(1999)}]{SM99}
\bibinfo{author}{\bibfnamefont{A.}~\bibnamefont{S{\"o}rensen}}
  \bibnamefont{and}
  \bibinfo{author}{\bibfnamefont{K.}~\bibnamefont{M{\o}lmer}},
  \bibinfo{journal}{Phys.\ Rev.\ Lett.} \textbf{\bibinfo{volume}{82}},
  \bibinfo{pages}{1971} (\bibinfo{year}{1999}).

\bibitem[{\citenamefont{Osnaghi et~al.}(2001)}]{OBA+01}
\bibinfo{author}{\bibfnamefont{S.}~\bibnamefont{Osnaghi}} \bibnamefont{et~al.},
  \bibinfo{journal}{Phys.\ Rev.\ Lett.} \textbf{\bibinfo{volume}{87}},
  \bibinfo{pages}{037902} (\bibinfo{year}{2001}).

\bibitem[{\citenamefont{Golub and van Loan}(1996)}]{GL96}
\bibinfo{author}{\bibfnamefont{G.~H.} \bibnamefont{Golub}} \bibnamefont{and}
  \bibinfo{author}{\bibfnamefont{C.~F.} \bibnamefont{van Loan}},
  \emph{\bibinfo{title}{Matrix Computations}} (\bibinfo{publisher}{Johns
  Hopkins University Press}, \bibinfo{year}{1996}), \bibinfo{edition}{3rd} ed.

\bibitem[{\citenamefont{Verstraete and Verschelde}(2003)}]{VV02}
\bibinfo{author}{\bibfnamefont{F.}~\bibnamefont{Verstraete}} \bibnamefont{and}
  \bibinfo{author}{\bibfnamefont{H.}~\bibnamefont{Verschelde}},
  \bibinfo{journal}{Phys.\ Rev.\ Lett.} \textbf{\bibinfo{volume}{90}},
  \bibinfo{pages}{097901} (\bibinfo{year}{2003}).

\bibitem[{\citenamefont{D{\"u}r et~al.}(1999)\citenamefont{D{\"u}r, Cirac, and
  Tarrach}}]{Dur99}
\bibinfo{author}{\bibfnamefont{W.}~\bibnamefont{D{\"u}r}},
  \bibinfo{author}{\bibfnamefont{J.~I.} \bibnamefont{Cirac}}, \bibnamefont{and}
  \bibinfo{author}{\bibfnamefont{R.}~\bibnamefont{Tarrach}},
  \bibinfo{journal}{Phys.\ Rev.\ Lett.} \textbf{\bibinfo{volume}{83}},
  \bibinfo{pages}{3562} (\bibinfo{year}{1999}).

\bibitem[{\citenamefont{Peres}(1996)}]{Per96}
\bibinfo{author}{\bibfnamefont{A.}~\bibnamefont{Peres}},
  \bibinfo{journal}{Phys.\ Rev.\ Lett.} \textbf{\bibinfo{volume}{77}},
  \bibinfo{pages}{1413} (\bibinfo{year}{1996}).

\bibitem[{\citenamefont{Horodecki et~al.}(1997)\citenamefont{Horodecki,
  Horodecki, and Horodecki}}]{HHH97}
\bibinfo{author}{\bibfnamefont{M.}~\bibnamefont{Horodecki}},
  \bibinfo{author}{\bibfnamefont{P.}~\bibnamefont{Horodecki}},
  \bibnamefont{and}
  \bibinfo{author}{\bibfnamefont{R.}~\bibnamefont{Horodecki}},
  \bibinfo{journal}{Phys.\ Rev.\ Lett.} \textbf{\bibinfo{volume}{78}},
  \bibinfo{pages}{574} (\bibinfo{year}{1997}).

\bibitem[{\citenamefont{Cirac et~al.}(2001)}]{CDK01}
\bibinfo{author}{\bibfnamefont{J.~I.} \bibnamefont{Cirac}}
  \bibnamefont{et~al.}, \bibinfo{journal}{Phys.\ Rev.\ Lett.}
  \textbf{\bibinfo{volume}{86}}, \bibinfo{pages}{544} (\bibinfo{year}{2001}).

\bibitem[{For(2000)}]{FortPhys}
\bibinfo{journal}{Fortschritte der Physik} \textbf{\bibinfo{volume}{48}},
  \bibinfo{pages}{issues 9 to 11} (\bibinfo{year}{2000}).

\end{thebibliography}

\end{document}